\documentclass[pre,twocolumn,preprintnumbers,amsmath,amssymb,superscriptaddress]{revtex4}
\usepackage[draft]{graphicx}
\usepackage{bm}
\usepackage[colorlinks=true, citecolor = blue]{hyperref}
\usepackage{lipsum}
\usepackage[dvipsnames]{xcolor}
\usepackage{soul}
\newcommand{\tc}{\textcolor{black}}

\newcommand{\as}{a_{\text{S}}}

\newcommand{\ns}{n_{\text{S}}}
\newcommand{\adagn}{a_{\text{S}}^{\dagger}n_{\text{S}}}
\newcommand{\na}{n_{\text{S}}a_{\text{S}}}

\usepackage{enumitem}

\begin{document}

\title{\boldmath Steady state quantum transport through an anharmonic oscillator strongly coupled to two heat reservoirs}

\author{Tianqi Chen}
\affiliation{Science, Mathematics and Technology Cluster, Singapore University of Technology and Design, 8 Somapah Road, 487372 Singapore}
\author{Vinitha Balachandran}
\affiliation{Science, Mathematics and Technology Cluster, Singapore University of Technology and Design, 8 Somapah Road, 487372 Singapore}
\author{Chu Guo}
\affiliation{Key Laboratory of Low-Dimensional Quantum Structures and Quantum Control of Ministry of Education, Department of Physics and Synergetic Innovation Center for Quantum Effects and Applications, Hunan Normal University, Changsha 410081, China}

\author{Dario Poletti}

\affiliation{Science, Mathematics and Technology Cluster, Singapore University of Technology and Design, 8 Somapah Road, 487372 Singapore}

\begin{abstract}
We investigate the transport properties of an anharmonic oscillator, modeled by a single-site Bose-Hubbard model, coupled to two different thermal baths using the numerically exact thermofield based chain-mapping matrix product states (TCMPS) approach. We compare the effectiveness of TCMPS to probe the nonequilibrium dynamics of strongly interacting system irrespective of the system-bath coupling against the global master equation approach in Gorini-Kossakowski-Sudarshan-Lindblad form. We discuss the effect of on-site interactions, temperature bias as well as the system-bath couplings on the steady state transport properties. Last we also show evidence of non-Markovian dynamics by studying the non-monotonicity of the time evolution of the trace distance between two different initial states.
\end{abstract}

\maketitle

\section{Introduction}
\label{sec:introduction}
Understanding heat flow in nanoscopic systems connected to thermal baths is interesting both for fundamental and practical reasons \cite{Ventra2011,Dhar2008,Whitney2017}. 
Of particular importance is the study of transport through extended strongly interacting system, as the presence of interactions can result in different phases of matter \cite{ProsenIlievski2011, GuoPoletti2016}, or ways to control transport \cite{GuoPoletti2015}, even resulting in strong or ideal current rectifiers \cite{BalachandranPoletti2018, BalachandranPoletti2019, LeePoletti2020}. 

Commonly used approaches to study heat transport in interacting systems are based on the
global master equation in Gorini-Kossakowski-Sudarshan-Lindblad form \cite{GoriniSudarshan1976, Lindblad1976, OQSBook} (from now on, we will refer to this approach as GME), or 
the Redfield master equation \cite{Redfield, AlonsoDeVegaReview}. However, these methods have two main difficulties: They can only be used for weak couplings between the system and the bath (especially the GME approach), and they typically require a full diagonalization of the Hamiltonian, which is computationally extremely difficult for larger open quantum systems, such as a chain of $14$ spins. 
More recently, in Ref.~\cite{XuPoletti2019}, some of us introduced a way to use the Redfield master equation without diagonalizing the Hamiltonian, and, thus, allowing one to study a spin chain of $20$ spins coupled to a bath.
Other available methods, such as Refs.~\cite{Hartree,Montecarlo,Segal,HEOM} and the Keldysh formalism \cite{Schwinger, Kadanoff, Keldysh, WangAgarwalla2014} can include non-perturbative effects and can capture non-Markovian dynamics, but can be limited in their applicability.   

Recently in Ref.~\cite{de2015thermofield}, the authors introduced a different approach based on three steps: (i) a thermal bath is mapped to two zero-temperature baths via a {\it thermofield} transformation, (ii) the collection of independent modes which forms a bath is mapped to a {\it chain} and (iii) the system plus the chains forming the bath are studied using a (MPS) algorithm \cite{White1992, White1993, Schollwock2011}, the method of choice to study the dynamics of one-dimensional strongly interacting quantum systems. We refer to this method, which has been studied and used in Refs.~\cite{de2015thermofield, de2015discretize, guo2018stable, XuPoletti2019, MascarenhasVega2017, DelftArrigoni2018,DelftWeichselbaum18}, as thermofield-based chain-mapping matrix product states (TCMPS). This method can allow the study of the dynamics of an interacting system coupled to a thermal bath exactly for any system-bath coupling strength. The only limitation of this method is due to the finite size of the chain used to model the bath, which implies a limited time for the accurate description of the system dynamics. Nonetheless, when the bath is modeled by long enough chains, it is still possible, as we show later, to study the steady-state properties of an interacting system coupled to a bath. We note that this method has strong analogies with time-evolving density matrix using orthogonal polynomials algorithm\cite{prior2010efficient, Plenio2010exactmapping, ChinPlenio2011, Plenio2011bookchapter, ChinPlenio2013, PriorPlenio2013, Plenio2014chainrepresentation, WoodsPlenio2015, RosenbachPlenio2016, TamascelliPlenio2019, NuesselerPlenio2019}.    
Although the long-term goal of this line of research would be the study of transport through a large interacting quantum system, here, we do the first steps by studying an anharmonic quantum oscillator (weakly and strongly) coupled to two baths at different temperatures. 
We point out that in the weak coupling regime, the heat transport through a quantum anharmonic oscillator has been recently studied in Refs.~\cite{ArchakManas2016, ChenSun2018}, however, our approach allows to study the system when strongly coupled to thermal baths. 
% \tc{For example, this will assist us to characterize quantum thermodynamics and control non-equilibrium heat transfer in strongly coupled single-site quantum systems that can be experimentally realized in superconducting circuit\cite{Singer2016,Pekola2018,Pekola2020} and cavity quantum electrodynamics (cQED)\cite{Schoelkopf2004,Rempe2014,Warburton2019}, which is essential to quantum devices at nanoscale.}
\tc{The use of TCMPS can assist in the study of quantum thermodynamics and heat transfer in electronic circuits, such as quantum dots, single-electron boxes, and superconducting qubits \cite{Pekola2015}, which are attractive candidates for the future development of quantum devices on the nanoscale.}

The reminder of this article is arranged as follows: In Sec.~\ref{sec:methods}, we introduce our setup of the anharmonic oscillator coupled to two reservoirs, and briefly review the concepts necessary to understand the TCMPS method. In Sec.~\ref{sec:results}, we study the non-equilibrium dynamics of the system using TCMPS, and compare it to GME. We, then, study the effect of interactions, temperatures, couplings between the system, and the bath on the steady-state transport properties of the system. Last, we discuss the emergence of non-Markovian dynamics. 
We draw our conclusions in Sec.~\ref{sec:conclusions}. Some of the mathematical details of the derivations of the formulas used, and ulterior numerical evidences, are given in the Appendices.

\section{Model and methods}
\label{sec:methods}
\subsection{Setup}
Our model consists of a single-site Bose-Hubbard system coupled to two thermal baths of harmonic oscillators at temperatures $T_L$ and $T_R$ as shown in Fig.~\ref{fig:startochain}(a). The total Hamiltonian can be written as
\begin{eqnarray}
\label{eq:singlesitebhm}
 \mathcal{\hat{H}} &=& \mathcal{\hat{H}}^{}_{\text{S}} + \sum_{\nu=L,R} \mathcal{\hat{H}}_{\text{B}}^{\nu} +\mathcal{\hat{H}}_{\text{I}}^{\nu}, \\ \nonumber
 \mathcal{\hat{H}}_{\text{S}} &=& \frac{U}{2} n_{\text{S}} \left( n_{\text{S}} - 1 \right) + \mu n_{\text{S}}, \\ \nonumber
 \mathcal{\hat{H}}_{\text{B}}^{\nu}&=&\int d\omega \, \omega b_{\omega}^{\nu \dagger}b_{\omega}^{\nu}, \left(\nu=L,R\right) \\ \nonumber
 \mathcal{\hat{H}}_{\text{I}}^{\nu} &=& \int d\omega \sqrt{\mathcal{J(\omega)}} \left(a_{\text{S}}^{\dagger}b_{\omega}^{\nu}+a_{\text{S}}b_{\omega}^{\nu \dagger} \right),
\end{eqnarray}
where $a_{\text{S}}^{\dagger}$ $(a_{\text{S}})$ is the creation (annihilation) operator of a boson in the anharmonic trap, $n_{\text{S}}=a_{\text{S}}^{\dagger}a_{\text{S}}$ counts the number of the particles in the system, $U$ is an on-site interaction which makes the system anharmonic, and $\mu$ is a local potential. $b_{\omega}^{\nu \dagger}$ (respectively, $b_{\omega}^{\nu}$) is the creation (annihilation) operator of each bath mode where $\nu$ is the index for left $(L)$ and right $(R)$ baths. $\mathcal{J}(\omega)$ is the spectral density of the baths, where we consider the Ohmic one $\mathcal{J}(\omega)=\gamma \omega$, and $\gamma$ is a dimensionless coupling constant. We consider baths which are initially prepared in a thermal state $\hat{\rho}_{\nu}={e^{-\beta^{\nu} \mathcal{\hat{H}}_{\text{B}}^{\nu}}}/{\mathrm{Tr}[  e^{-\beta ^{\nu} \mathcal{\hat{H}}_{\text{B}}^{\nu}} ]}$, with $\beta^{\nu}$ being the inverse temperature $\beta^{\nu}=1/k_BT_{\nu}$ where $k_B$ is the Boltzmann constant.

In our calculations, the original bath Hamiltonian $\mathcal{\hat{H}}_{\text{B}}$ is discretized into $N$ oscillators as
\begin{align}
	\mathcal{\hat{H}}^{\nu}_{\text{BD}} &= \sum_{k=1}^{N}\omega_k b_k^{\nu \dagger}b^{\nu}_k, \label{eq:bath_discrete}
\end{align}
where $\omega_k=k\Delta \omega$ with $\Delta \omega=\omega_{\text{c}}/N$ where $\omega_{\text{c}}$  is the frequency cut-off of the bath such that $\mathcal{J}(\omega)=0$ for $\omega > \omega_{\text{c}}$. The interaction Hamiltonian, thus, needs to be rewritten, and it becomes
\begin{align}
\label{eq:interactionHamiltonian}
	\mathcal{\hat{H}}^{\nu}_{\text{ID}} &=\sum_{k=1}^{N}\sqrt{\mathcal{J}_k}\left( a_{\mathrm{S}}^{\dagger}b^{\nu}_k + a_{\mathrm{S}} b_k^{\nu \dagger}\right),
\end{align}
with $\mathcal{J}_k=\int_{\omega_k}^{\omega_{k+1}}d\omega\,\mathcal{J}(\omega)\approx\mathcal{J}(\omega_k)\Delta \omega$.

\subsection{Thermofield based Chain-mapping Method}
\begin{figure}[h]
\centering
\includegraphics[width=1.0\columnwidth,draft=false]{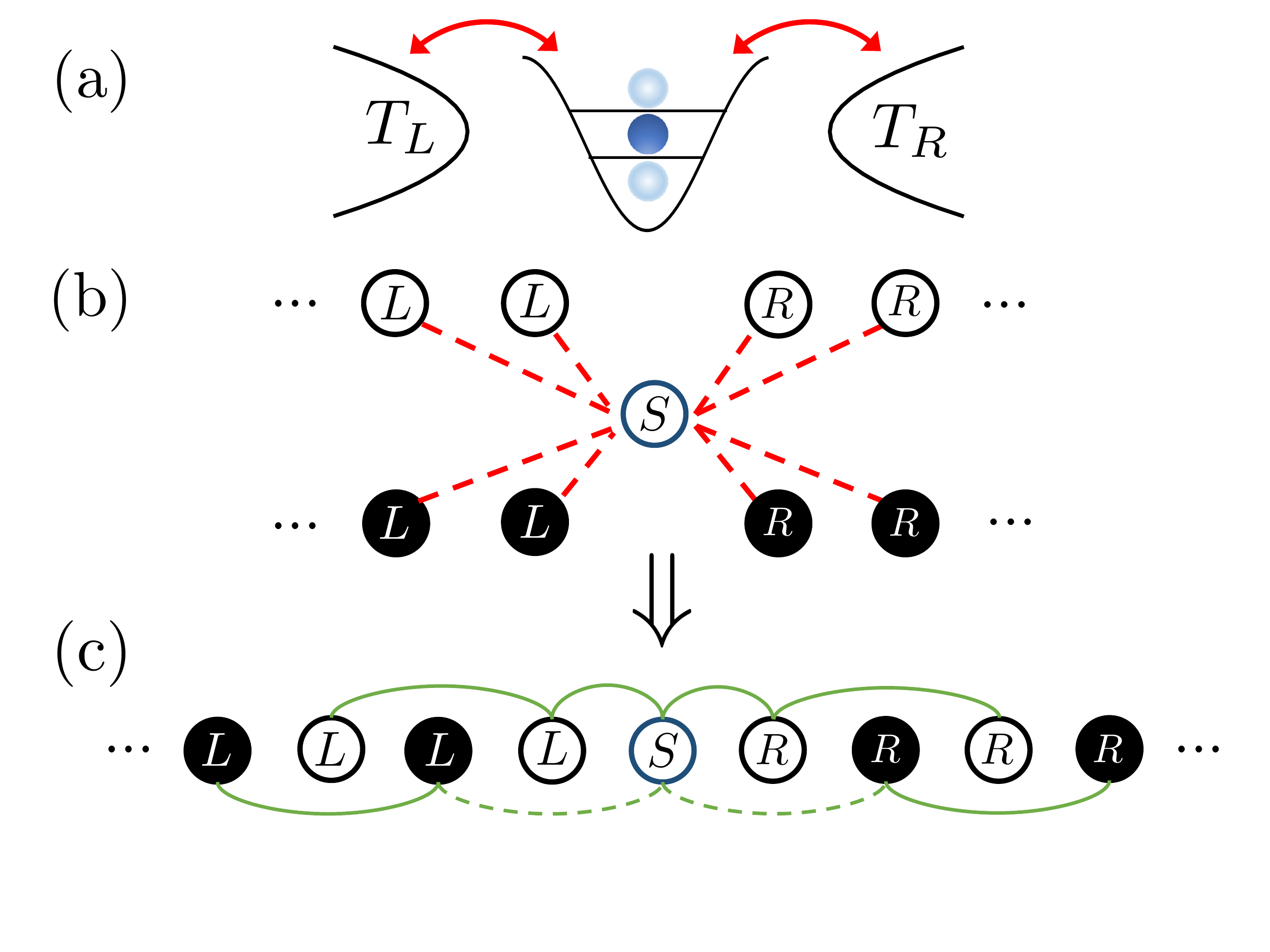}
\caption{\label{fig:startochain} (a) A single site Bose Hubbard model is coupled to two baths at different temperatures: left bath temperature is $T_L$ and the right bath temperature is $T_R$; (b) schematics of thermofield-based transformation: The bath on each side is discretized and is mapped to two baths at zero temperature labeled as solid and hollow circles. The system is coupled to all bath modes on each side. (c) Star-to-chain mapping: The total system is mapped to one single chain on each side with next-nearest neighbor tunnelings.}
\end{figure}

We, now, briefly review the thermofield plus star-to-chain mapping approach introduced in Ref.~\cite{de2015discretize} for studying the systems coupled to thermal baths and which can be implemented using MPSs. For clarity of explanation, in this section, we consider the system to be coupled to a single thermal bath, and, hence, the bath index $\nu$ is neglected. The two-bath case is a simple extension of this one. The thermofield approach consists of adding an auxiliary and decoupled bath to the original bath, and the new enlarged bath Hamiltonian becomes
\begin{align}
	\mathcal{\hat{H}}_{\text{B,C}} &= \mathcal{\hat{H}}_{\text{B}} + \mathcal{\hat{H}}_{\text{C}} = \sum_{k=1}^{N}\omega_k b_k^{\dagger}b_k - \sum_{k=1}^{N}\omega_k c_k^{\dagger}c_k,
\end{align}
where $c_k(c_k^{\dagger})$ are annihilation (creation) operators for the auxiliary bath [see Fig.~\ref{fig:startochain}(b) for a depiction of the case with two baths]. We,then, apply the following thermal Bogoliubov transformation,
\begin{align}
\label{eq:Bogliubovtransformation}
	a_{1,k} &= e^{-\mathrm{i}G} b_k e^{\mathrm{i} G} = \cosh(\theta_k)b_k-\sinh(\theta_k)c_k^{\dagger}, \\ \nonumber
	a_{2,k} &= e^{-\mathrm{i}G} c_k e^{\mathrm{i} G} = \cosh(\theta_k)c_k-\sinh(\theta_k)b_k^{\dagger},
\end{align}
where $G=\mathrm{i}\sum_k \theta_k(b_k^{\dagger}c_k^{\dagger}-c_kb_k)$ with $\cosh{(\theta_k)}=\sqrt{1+n_k},\sinh{(\theta_k)}=\sqrt{n_k}$ and $n_k=1/(e^{\beta \omega_k}-1)$ is the number of excitations in mode $k$. Thus, the total Hamiltonian of the system plus bath after thermofield transformation becomes
\begin{align}
	\label{eq:spbhamiltonian}
	\mathcal{\hat{H}}^{\text{TF}} &= \mathcal{\hat{H}}_{\text{S}} + \mathcal{\hat{H}}_{\text{B,C}}^{\text{TF}} + \mathcal{\hat{H}}_{\text{ID}}^{\text{TF}} \\ \nonumber &= \mathcal{\hat{H}}_{\text{S}} + \sum_{k=1}^{N}\omega_k(a_{1,k}^{\dagger}a_{1,k}-a_{2,k}^{\dagger}a_{2,k}) \\ \nonumber &+ \sum_{k=1}^{N}g_{1,k}\left(a_{\mathrm{S}}^{\dagger}a_{1,k}+a_{\mathrm{S}} a_{1,k}^{\dagger}\right)
	\\ \nonumber &+ \sum_{k=1}^{N}g_{2,k}\left(a_{\mathrm{S}} a_{2,k} + a_{\mathrm{S}}^{\dagger}a_{2,k}^{\dagger} \right),
\end{align}
where $g_{1,k}$ and $g_{2,k}$ are new coupling coefficients and $g_{1,k}=\sqrt{\mathcal{J}_k}\cosh{(\theta_k)}$, $g_{2,k}=\sqrt{\mathcal{J}_k}\sinh{(\theta_k)}$. 
At this point, the system is coupled with all the baths modes (also known as a star configuration). Whereas MPSs are the method of choice to study the physics of one-dimensional strongly interacting systems, the star configuration is typically not ideal for matrix product state calculations as it requires long-range couplings \cite{footnote_on_fermions}. To circumvent this problem, the star geometry in Eq.~\eqref{eq:spbhamiltonian} can be transformed into short-range ones (`chain' geometry) which is suitable for MPS time evolution by performing a star-to-chain mapping \cite{de2015thermofield, de2015discretize, prior2010efficient, Plenio2010exactmapping, ChinPlenio2011, Plenio2011bookchapter, ChinPlenio2013, PriorPlenio2013, Plenio2014chainrepresentation, WoodsPlenio2015, RosenbachPlenio2016, TamascelliPlenio2019, NuesselerPlenio2019}. This can be implemented, for instance, by a Lanczos tridiagonalization which gives new orthogonal basis to represent the baths (real and auxiliary) \cite{Gautschi2005OP, de2015discretize} with modes $d_{\zeta,j}$ $(d_{\zeta,j}^{\dagger})$ which are coupled along a chain and of which only one is coupled to the system.
As a result, the initial thermal bath in Eq.~(\ref{eq:bath_discrete}) becomes two decoupled tight-binding chains with nearest-neighbor tunneling coefficients $\beta_{\zeta,j}$, and the on-site potential as $\alpha_{\zeta,j}$ where $\zeta=1,2$ and $j$ is a discrete index for the site numbers [see a single side bath chain in Fig.~\ref{fig:startochain}(c)]. More specifically, the total Hamiltonian, then, becomes
\begin{align}
\label{eq:star-to-chainhamiltonian}
	& \mathcal{\hat{H}}^{\text{TCMPS}} = \mathcal{\hat{H}}_{\text{S}} + \sum_{\zeta} \left[  \sum_{j=1}^{N_{\text{chain}}}\alpha_{\zeta,j} d^{\dagger}_{\zeta,j}d_{\zeta,j} \right. \\
\nonumber &\left. + \sum_{j=1}^{N_{\text{chain}}-1}\beta_{\zeta,j}\left(d^{\dagger}_{\zeta,j}d_{\zeta,j+1}+d_{\zeta,j}d^{\dagger}_{\zeta,j+1} \right) \right] \\ \nonumber &+\left[\beta_{1,0} \left(a_{S}^{\dagger} d_{1,1} + a^{}_S d^{\dagger}_{1,1} \right)+\beta_{2,0} \left(a_{S}^{\dagger} d^{\dagger}_{2,1} + a^{}_S d_{2,1} \right)\right].
\end{align}
where $N_{\text{chain}}$ is the length of the bath in the `chain' geometry. For our two-bath setup in Eq.~(\ref{eq:singlesitebhm}), the baths on both sides undergo the same thermofield-based transformation and star-to-chain mapping. This conclusively results in two bath chains aligned on both sides in one dimension where each mode in each bath chain is coupled to its next-to-nearest neighbor [see Fig.~\ref{fig:startochain}].

The final form of the total Hamiltonian then becomes
\begin{align}
\label{eq:totaldisHam}
&\mathcal{\hat{H}}^{\text{TCMPS}}_{\text{tot}}= \mathcal{\hat{H}_{\text{S}}} +\mathcal{\hat{H}_{\text{BD}}}^L+\mathcal{\hat{H}_{\text{BD}}}^R+ \mathcal{\hat{H}_{\text{ID}}}^L + \mathcal{\hat{H}_{\text{ID}}}^R  \\ \nonumber &= \frac{U}{2} n_{\text{S}} \left( n_{\text{S}} - 1 \right) + \mu n_{\text{S}} + \sum_{\zeta,\nu} \left[  \sum_{j=1}^{N_{\text{chain}}}\alpha^{\nu}_{\zeta,j} d^{\nu\dagger}_{\zeta,j}d^{\nu}_{\zeta,j} \right. \\
\nonumber &\left. + \sum_{j=1}^{N_{\text{chain}}-1}\beta^{\nu}_{\zeta,j}\left(d^{\nu\dagger}_{\zeta,j}d^{\nu}_{\zeta,j+1}+d^{\nu}_{\zeta,j}d^{\nu\dagger}_{\zeta,j+1} \right) \right] \\ \nonumber &+\sum_{\nu}\left[\beta^{\nu}_{1,0} \left(a_{S}^{\dagger} d^{\nu}_{1,1} + a^{}_S d^{\nu\dagger}_{1,1} \right)+\beta^{\nu}_{2,0} \left(a_{S}^{\dagger} d^{\nu \dagger}_{2,1} + a^{}_S d^{\nu}_{2,1} \right)\right].
\end{align}
We point out, here, that one could have simply implemented the star-to-chain mapping on Eq.~(\ref{eq:bath_discrete}) and (\ref{eq:interactionHamiltonian}) and could already have studied the problem with matrix product states. However, in this case, the modes would have a non-zero occupation, and one would potentially have to keep a large number of possible occupations of the new modes. Instead, an important advantage of using the thermofield transformation is that the resulting new modes $d_{\zeta,j}$ are at zero temperature, and, hence, they are empty.

We also stress that the thermofield plus star-to-chain mapping is an exact representation of the system plus (discretized) baths for any coupling strength. However, given the finite number of modes considered for the baths, this representation is valid only for a finite time which increases with the length of the chain. As we will show later, we will consider long-enough chains so as to reach a steady state.  

\subsection{Initial condition} 
In the following calculations, unless specified otherwise, we consider the initial condition 
\begin{align}
| \psi(0)\rangle = | 0\rangle_S \otimes | 0 \rangle_{L_1}  \otimes | 0 \rangle_{L_2} \otimes  | 0\rangle_{R_1}  \otimes   | 0 \rangle_{R_2}       
\end{align} 
i.e. a tensor product of the vacuum for the operators ${a}_S$, ${d}^L_{1,j}$, ${d}^L_{2,j}$, ${d}^R_{1,j}$ and ${d}^R_{2,j}$, which corresponds to simulate the initial condition $\hat{\rho}(0) = |0\rangle  \langle 0 | \otimes \hat{\rho}_L \otimes \hat{\rho}_R$.

\subsection{MPSs}
Here, we summarize the main concepts of the MPS numerical approach. MPS is fundamentally a variational ansatz in which a vector $\vert \psi \rangle$ (which effectively can represent a wave-function or a density matrix) is described by a product of matrices or, more generally, tensors. The vector $\vert \psi \rangle$ which, here, represents a wave-function over $M$ sites, can be written in a basis $\vert \psi \rangle = \sum_{a_1,\dots a_M} c_{a_1,\dots a_M} \vert a_1,\dots a_M \rangle$ where the local \tc{index} $a_j$ can take a certain number of values which we call the local Hilbert space $d$.
The MPS variational ansatz can, thus, be written as
\begin{align}
\vert \psi \rangle = \sum_{a_1,\dots a_M} \otimes_{j=1}^M W^{a_j}_{\alpha_{j-1},\alpha_{j}} \vert a_1,\dots a_M \rangle,  \label{eq:mps}
\end{align}
where \tc{$W^{a_j}_{\alpha_{j-1},\alpha_{j}}$ is a rank-$3$ tensor with the $a_j$ labeling the local quantum state at site $j$ from the possible state of the local physical Hilbert space, and} the $\alpha_j$'s are auxiliary indices which we take to be $D$ at most, the so called {\it bond dimension}. In Eq.~(\ref{eq:mps}), the tensor contraction over repeated indices $\alpha_j$ is implied, and, naturally $\alpha_0=\alpha_M=1$.

\section{Results}
\label{sec:results}

In this section, we study the steady-state properties of the anharmonic oscillator in Eq.~(\ref{eq:singlesitebhm}) using TCMPS. We perform the unitary time evolution of the system and baths using the Hamiltonian in Eq.~(\ref{eq:totaldisHam}) with MPS by implementing a non-number conserving second-order Suzuki-Trotter algorithm with swap gates \cite{stoudenmire2010minimally,Suzuki1990}. The evolution time step is chosen to be $0.04\hbar/\mu$, and  convergences of the simulations are confirmed by checking the truncation errors after repeating the runs for different values of the maximum bond dimension, the local Hilbert space dimension, and the length of bath chains. We find that keeping a maximum number of levels for the modes $d^\nu_{\zeta,j}$ equal to five, a maximum bond dimension of $300$ auxiliary levels, and a maximum chain length $L=50$, allows us to produce precise simulations, with errors on the observables of, at most, $10^{-5}$.  Throughout this paper, we have worked in units such that $\hbar=\mu=1$.

\subsection{Occupation versus time}
\label{subsec:capability}

\begin{figure}[ht]
\centering
\includegraphics[width=1.0\columnwidth, draft=false]{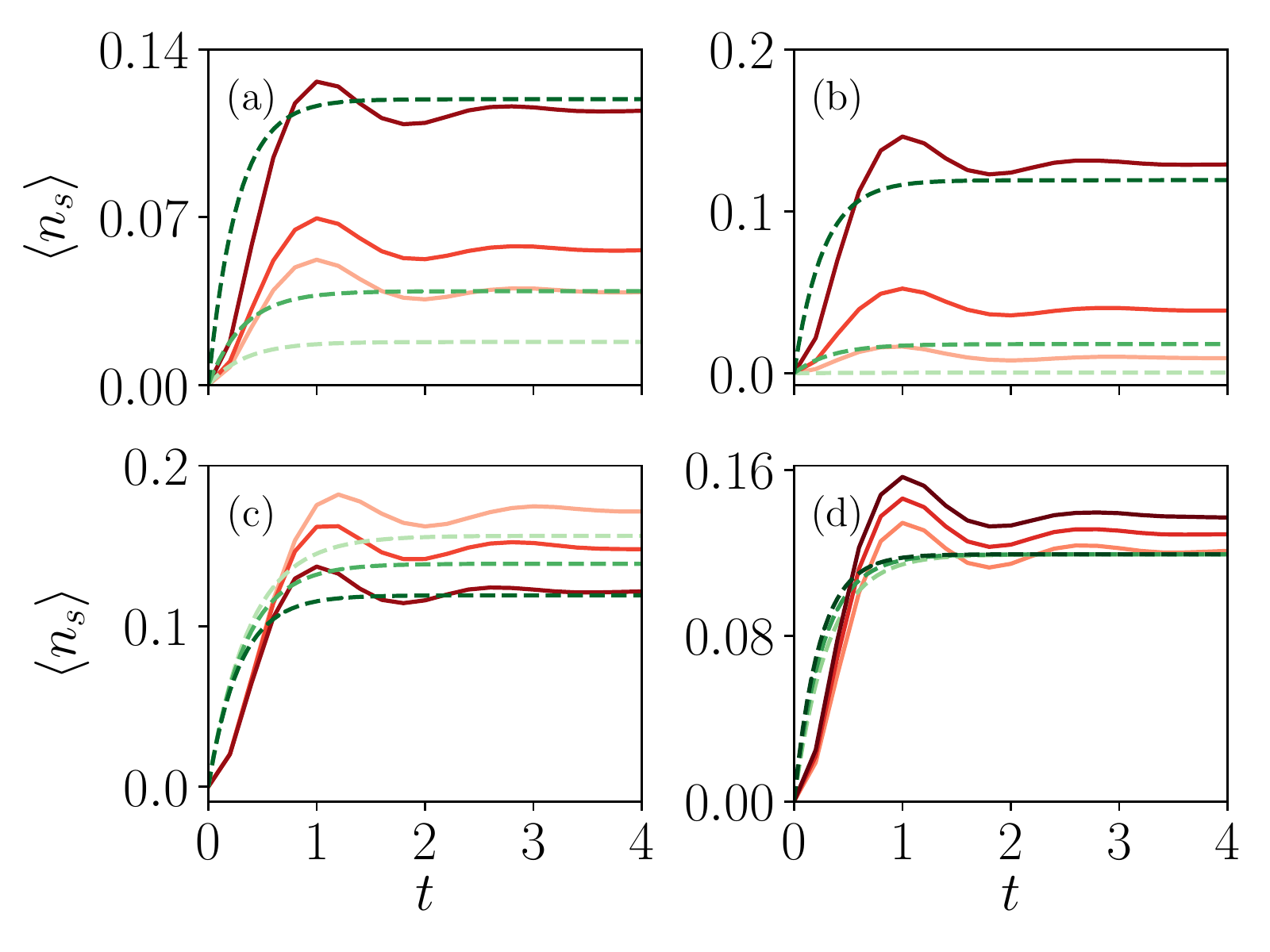}
\caption{\label{fig:timedynamics} Time evolution of system's occupation $\langle {n}_S\rangle$ for (a) different ratios of temperature biases $T_L/T_R=2$, $2.5$ and $4.0$, where $T_R=0.125$. Lighter to darker colors represent the ratios $T_L/T_R$ from low to high. (b) Time evolution for low, intermediate and high temperature regimes with the same temperature bias $T_L/T_R=2$: $T_L=0.125, T_R=0.0625$, $T_L=0.25, T_R=0.125$ and $T_L=0.5, T_R=0.25$. Lighter to darker colors represent the temperatures of both baths from low to high. Other parameters for (a) and (b) are $U=1.5$ and $\gamma=0.1225$. (c) Time evolution for low to high system interactions: $U=0$, $U=0.25$, and $U=2.25$ with the coupling  $\gamma=0.105625$, $T_L=0.5$ and $T_R=0.25$. Lighter to darker colors represent $U$ from low to high. (d) Time evolution for different system-bath coupling constants $\gamma=0.09, 0.1225, 0.16$ for an interaction $U=1.5$, $T_L=0.5$ and $T_R=0.25$. Lighter to darker colors represent the system-bath coupling constants from low to high. Dotted lines correspond to results obtained using the global master equation, see Appendix \ref{app:GKSL}.}
\end{figure}

We first consider the average occupation of the anharmonic oscillator $\langle n_{\text{S}}\rangle$ versus time for different bath temperatures, interaction strengths, and system-bath coupling magnitudes. The results from TCMPS calculations are depicted by continuous lines in Fig.~\ref{fig:timedynamics}. For all the parameters considered the average occupation reaches a steady value, indicating that we have considered long enough chains to represent the baths.
In Fig.~\ref{fig:timedynamics}(a), we plot $\langle n_{\text{S}}\rangle$ versus time for different temperature ratios between the hot bath on the left ($T_L=1/\beta_L$) and the cold bath on the right ($T_R=1/\beta_R$), whereas keeping $T_R=0.125$ (darker lines correspond to larger ratios). What we observe is that, for larger $T_L$, the occupation increases.
We note, here, that the evolution time is sufficient to reach the steady state even starting from quite different initial conditions as shown in Appendix.~\ref{app:steadystateconvergence}.
A similar physical insight is gained from Fig.~\ref{fig:timedynamics}(b), in which the ratio $T_L/T_R=2$ is kept constant, but $T_L$ is varied from $0.125$ to $0.5$.

In Fig.~\ref{fig:timedynamics}(c), we investigate the effect of the interaction $U$ on the system occupation for high temperature bias. The initial dynamics of the occupation is independent of the system interaction. However, the steady-state value of the occupations is lower in the presence of stronger interactions as the occupations of higher levels are suppressed. Indeed, for strong interactions, the system and the temperatures considered, the anharmonic trap could be well approximated by a two-level system.
In Fig.~\ref{fig:timedynamics}(d), we study the dependence on the system-bath coupling. Our calculations show that a longer time is required to reach the steady state for weaker couplings. From a computational point of view, we point out that we cannot use too large couplings because they require a larger number of local levels, auxiliary levels in the MPS code, and longer chains. Hence, we restrict our analysis to couplings  between $0.04$ and $0.16$. It is clear from the figure that different coupling strengths lead to different steady-state occupations and, hence, different steady states. This is because the system and the bath are more strongly coupled and correlated. 

In all panels of Fig.~\ref{fig:timedynamics}, we use green dashed lines to indicate results from a global master equation (GME) description of the system dynamics in Gorini-Kossakowski-Sudarshan-Lindblad form \cite{GoriniSudarshan1976, Lindblad1976, OQSBook}. Such a description is most accurate for very weak couplings between the system and the bath, and at higher temperatures (see Appendix.~\ref{app:GKSL} for the relevant equations), which is, indeed, what we observe in Fig.~\ref{fig:timedynamics} when comparing the results from the GME to those from the exact description via TCMPS. In particular, we note that even when the steady-state value of $\langle n_{\text{S}}\rangle$ from GME is similar to that from TCMPS, the time evolution can be significantly different as, for instance, the GME has no oscillations. Another important point is that, whereas the steady-state value is expected to be different when the coupling is not very weak, the GME predicts a steady-state value which is independent of $\gamma$, see Fig.~\ref{fig:timedynamics}(d). For the same reason, in Fig.~\ref{fig:timedynamics}(c), we observe that the GME and the TCMPS descriptions agree better at larger interaction $U$ because the coupling between the system and the bath is effectively weaker.

\subsection{Steady-state particle and energy currents}
We now shift our focus towards the steady-state properties of the system. It follows from the continuity equation
\begin{align}
	\frac{\partial \langle n_{\text{S}} \rangle}{\partial t} &= \left \langle {i}\left[ \mathcal{\hat{H}}, n_{\text{S}} \right]\right\rangle=-\left(J_{\text{p}}^{R} - J_{\text{p}}^L \right),
\end{align}
that the particle current $J_{\text{p}}^L$ ($J_{\text{p}}^R$) from the left (right) bath are given by
\begin{align}
	J_{\text{p}}^L &= 2\beta_{1,0}^L \operatorname{Im}\! \left(\langle a_{S}^{\dagger} d_{1,1}^L \rangle \right) + 2\beta_{2,0}^L \operatorname{Im}\! \left(\langle d_{2,1}^{L \dagger} a_{\text{S}}^{\dagger} \rangle \right),  \\
	J_{\text{p}}^R &= 2\beta_{1,0}^R \operatorname{Im}\! \left( \langle a_{\text{S}}d_{1,1}^{R\dagger} \rangle  \right) + 2\beta_{2,0}^R \operatorname{Im}\! \left( \langle a_{\text{S}}d_{2,1}^R \rangle \right),
\end{align}
where $\operatorname{Im}\left(\cdot\right)$ stands the for imaginary part.

Similarly, we can define the energy current as the rate of change of energy of the system,
\begin{align}
\label{eq:heatcurrentderivation}
\frac{\partial \left \langle\hat{\mathcal{H}}_{\text{S}}\right \rangle}{\partial t} &= \left \langle {i}\left [ \mathcal{\hat{H}},\hat{\mathcal{H}}_{\text{S}}  \right]\right\rangle = - \left( J_{\text{e}}^R - J_{\text{e}}^L\right),
\end{align}
and the energy current from the left, $J_{\text{e}}^L$, and the right, $J_{\text{e}}^R$, baths are given by
\begin{align}
\label{eq:leftheatcurrent}
J_{\text{e}}^L&=2U \beta_{1,0}^L
\mathrm{Im}\! \left( \langle d_{1,1}^L\adagn \rangle \right)  +2U \beta_{2,0}^L \mathrm{Im}\!\left( \langle d_{2,1}^{L\dagger} \adagn  \rangle \right)     \nonumber  \\
&+2\mu\left[\beta_{1,0}^L\mathrm{Im}\!\left( \langle d_{1,1}^L \as^{\dagger} \rangle \right)  +\beta_{2,0}^L \mathrm{Im} \! \left( \langle d_{2,1}^{L\dagger}\as^{\dagger} \rangle\right)\right]. \\ \nonumber
J_{\text{e}}^R&= 2U \beta_{1,0}^R \mathrm{Im} \! \left( \langle d_{1,1}^{R\dagger} \na \rangle \right) +2U\beta_{2,0}^R\mathrm{Im} \! \left( \langle d_{2,1}^R \na \rangle \right) \\
&+ 2\mu\left[\beta_{1,0}^R\mathrm{Im} \! \left( \langle d_{1,1}^{R\dagger}\as \rangle \right) +\beta_{2,0}^R \mathrm{Im} \! \left( \langle d_{2,1}^R \as \rangle \right) \right].
\end{align}
In Appendix.~\ref{app:heatcurrent} and \ref{app:particlecurrentderivation}, we show the detailed derivations of the above expressions.

While discussing energy current, it is worth opening parentheses on the physical meaning of the expressions used. As shown in Eq.~(\ref{eq:heatcurrentderivation}), the starting point is the energy of the system which is given by the system Hamiltonian $\hat{\mathcal{H}}_{\text{S}}$. It is, however, important to note that such a starting point only makes sense when the coupling between the system and the bath is small enough, otherwise, it would not be clear how to consider the energy of the system, as the energy due to the interaction with the baths may not be negligible. For the parameters considered, the ratio $|\langle\hat{\mathcal{H}}_{\text{I}} \rangle / \langle\hat{\mathcal{H}}_{\text{S}} \rangle|$ varies between $5\times 10^{-3}$ (e.g., for small $\gamma$, high temperatures, and large interaction $U$) and $1.7\times 10^{-2}$ (for large $\gamma$, low temperatures, and small interaction $U$). This implies that, for the parameters considered, it is fairly meaningful to use the concept of {\it energy of the system}. 
Studies which investigate heat current, and its definition, in the strong system-bath coupling regime can be found in Refs.~\cite{Campisi2009, Talkner2016, Esposito2015, KatzKosloff2016, Marti2018, Jarzynski, Strasberg2018, Sanchez2016, Carrega2016, Gelbwaser2015, He2018}. 

\begin{figure}[h]
\centering
\includegraphics[width=1.\columnwidth, draft=false]{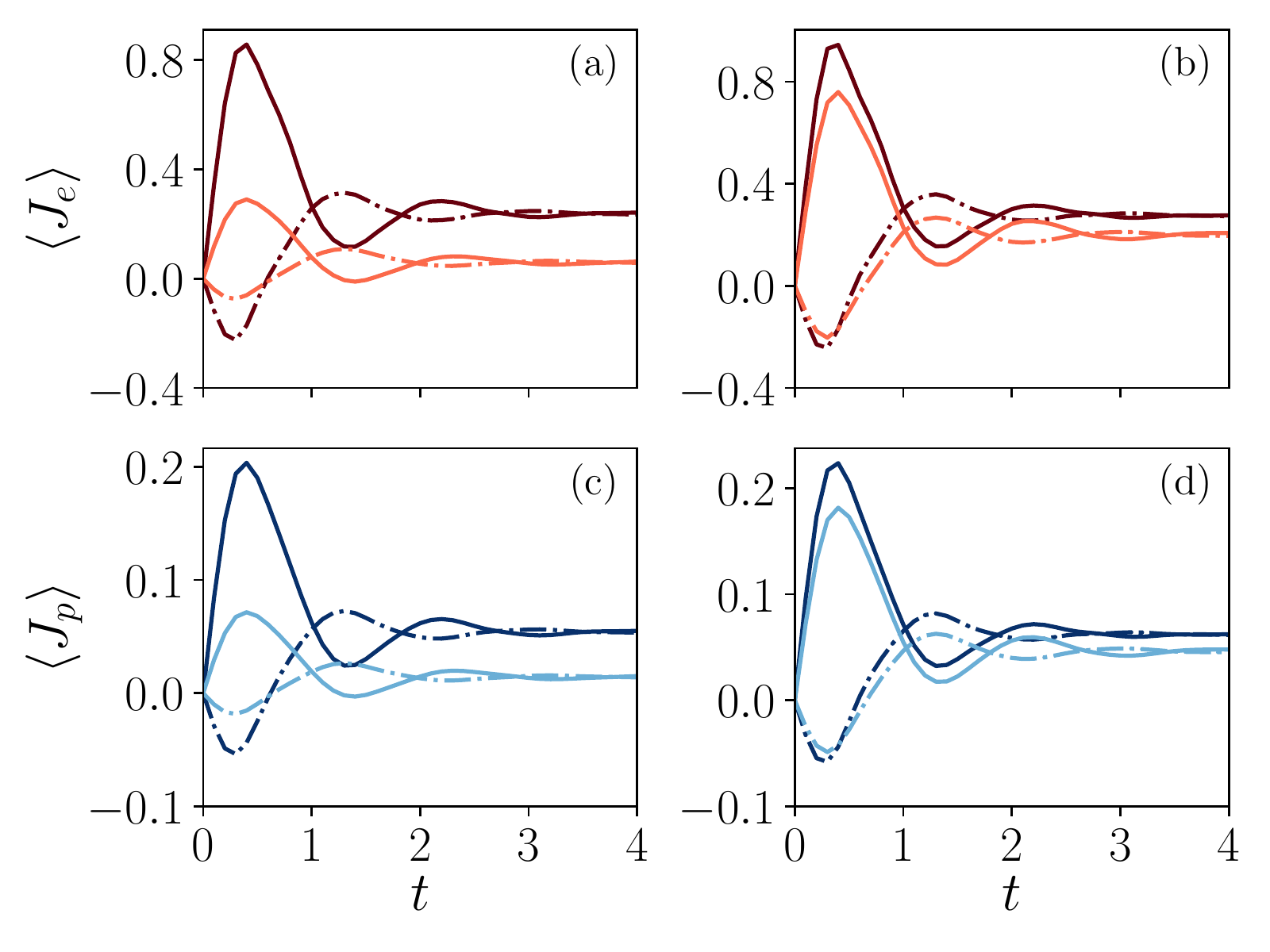}
\caption{\label{fig:heatparticlecurrenttimedynamics} \tc{Time evolution of energy current: (a) and (b), and particle current: (c) and (d). Left panels (a) and (c) are for low and high temperature regimes with the same temperature bias $T_L/T_R = 2$. Darker-colored curves are for $T_L=0.5$, and $T_R=0.25$, and lighter-colored curves are for $T_L=0.125$, and $T_R=0.0625$. The system-bath coupling strength is $\gamma=0.1225$. Right panels (b) and (d) are for different system-bath coupling strengths $\gamma$. Darker-colored curves are for $\gamma=0.16$, and lighter-colored curves are for $\gamma=0.09$. The bath temperature is as follows: $T_L=0.5$, and $T_R=0.25$. In all panels, solid lines are for currents from the left bath, and dot-dashed lines are for currents from the right bath. Other parameters used are as follows: $U=1.5$, and $\omega_{c}=2.5$.}}
\end{figure}     

\tc{In Fig.~\ref{fig:heatparticlecurrenttimedynamics}, we show the energy $\langle J_{\text{e}} \rangle$ and particle currents $\langle J_{\text{p}} \rangle$ from both left and right baths versus time for different system-bath coupling strengths and bath temperatures. It is observed that the energy and particle currents from the left and right baths reach the same steady state within the time considered, for both low and high temperatures [Figs.~\ref{fig:heatparticlecurrenttimedynamics}(a), and \ref{fig:heatparticlecurrenttimedynamics}(c)], as well as for small and large system-bath coupling strengths [Figs.~\ref{fig:heatparticlecurrenttimedynamics}(b), and \ref{fig:heatparticlecurrenttimedynamics} (d)]. %We also note that the oscilatory behavior of both energy and particle currents to reach the steady state is consistent.
  }

\begin{figure}[ht]
\centering
\includegraphics[width=1.0\columnwidth, draft=false]{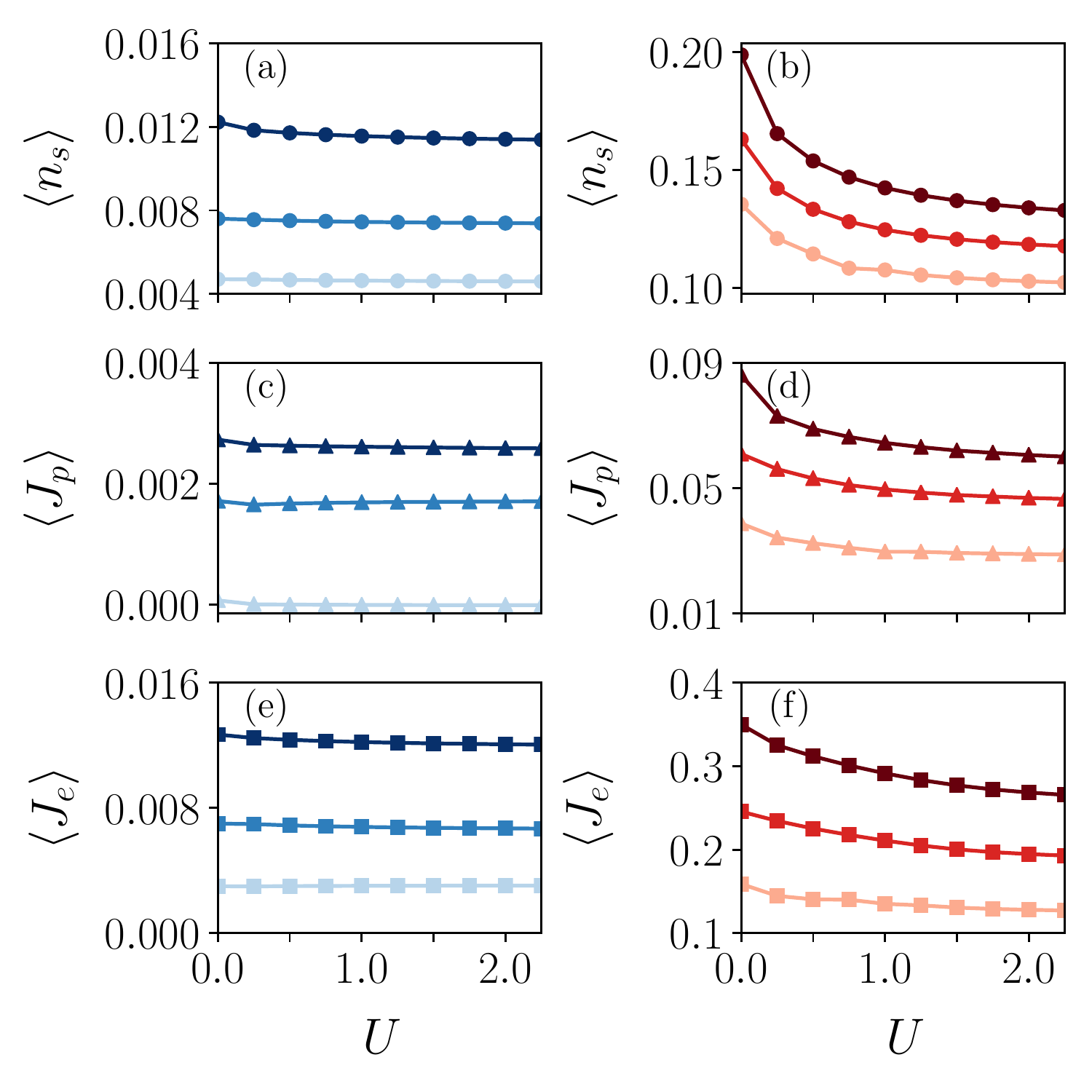}
\caption{\label{fig:interactioneffect}Effect of interaction $U$ on: (a) and, (b) system occupation, (c) and, (d) particle current, and (e) and, (f) energy current for different magnitudes of the system-bath coupling constant $\gamma$. Left panels (a), (c), and (e) are for low temperatures $T_L=0.125, T_R=0.0625$ and right panels (b), (d), and (f) for high temperatures $T_L=0.5, T_R=0.25$. Curves from lighter to darker color in all panels correspond to smaller to larger system-bath coupling strengths with values of $\gamma=0.04,0.09$ and $0.16$ respectively. Other parameters used are $\mu=1$, $\omega_c=2.5$.}
\end{figure}

In Fig.~\ref{fig:interactioneffect}, we study the average density [Figs.~\ref{fig:interactioneffect}(a), and \ref{fig:interactioneffect} (b)], particle current [Figs.~\ref{fig:interactioneffect}(c), and \ref{fig:interactioneffect} (d)] and energy current [Figs.~\ref{fig:interactioneffect}(e), and \ref{fig:interactioneffect}(f)], for lower [Figs.~\ref{fig:interactioneffect}(a), \ref{fig:interactioneffect}(c), and \ref{fig:interactioneffect}(e)] and higher [Figs.~\ref{fig:interactioneffect}(b), \ref{fig:interactioneffect}(d), and \ref{fig:interactioneffect}(f)] temperatures. In each panel, different curves represent different system-bath couplings. It is only for higher temperatures, for which there is larger occupation of the anharmonic oscillator, that a clear effect of the interactions becomes more apparent. In particular, Fig.~\ref{fig:interactioneffect} shows that stronger interactions result in lower occupation, particle, and energy currents.

\begin{figure}[h]
\centering
\includegraphics[width=1.0\columnwidth, draft=false]{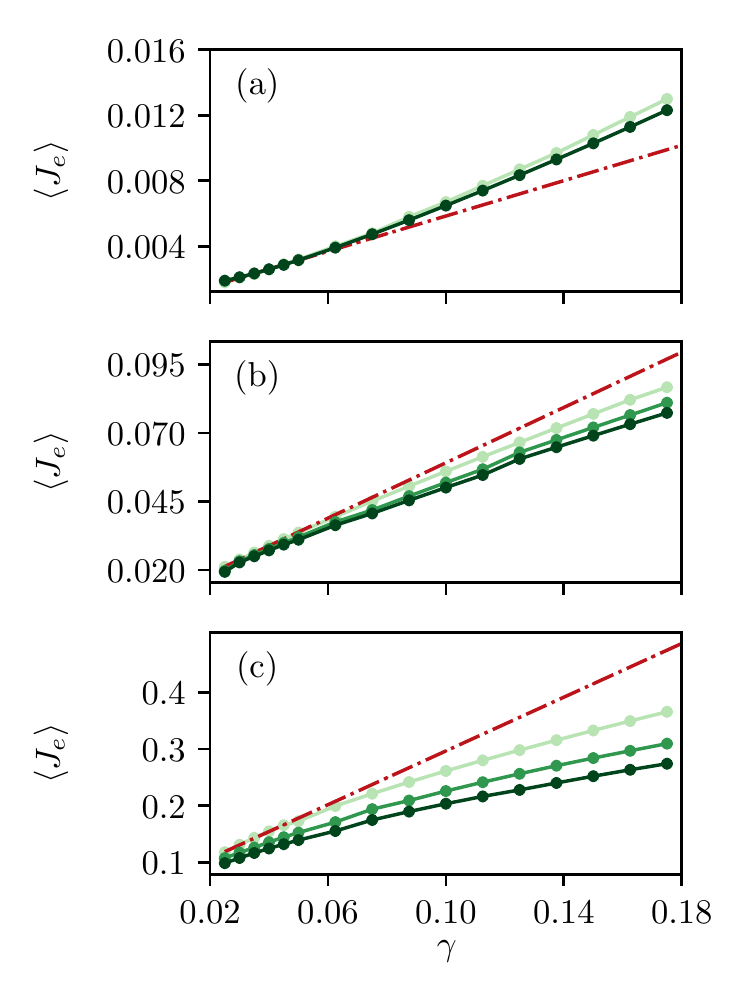}
\caption{\label{fig:heatcurrenttemperature}Steady-state energy current as a function of system-bath coupling strength $\gamma$ for: (a) low temperatures $(T_L=0.125, T_R=0.0625)$, (b) intermediate temperatures $(T_L=0.25, T_R=0.125)$, and (c) high temperatures $(T_L=0.5, T_R=0.25)$. The ratio of bath temperature bias is the same $(T_L/T_R=2)$ for all panels. In each panel, solid curves from lighter to darker green color indicate smaller to larger strength of interaction: $U = 0, 0.75$ and $2.25$. The red dot-dashed lines indicate a linear fit for $U=0$ and weak system-bath coupling $\gamma < 0.04$. In all panels we used $\omega_c=2.5$.}
\end{figure}

In Fig.~\ref{fig:heatcurrenttemperature}, we study the effect of the system-bath coupling $\gamma$ on the steady-state energy current for different values of the interaction $U$, and in different temperature regimes. For lower temperatures, Fig.~\ref{fig:heatcurrenttemperature}(a), we observe super-linear dependence of the current with $\gamma$, whereas at intermediate and higher temperatures, the dependence is sublinear. We associate the super-linear dependence at low temperatures with the relevance, in that temperature regime\cite{cotunnelingfootnote}, of a coherent two-bosons process known as ``cotunneling'' \cite{Ruokola, Segal2, WuSegal2010}.

\subsection{Signatures of non-Markovianity}
\label{subsec:nonMarkovian}

Strong system-bath couplings typically lead to non-Markovian dynamics. Hence, in this section, we analyze the non-Markovianity of the dynamics by measuring the trace distance between two quantum states $\hat{\rho}_1$ and $\hat{\rho}_2$ as

\begin{align}
D(\hat{\rho}_1,\hat{\rho}_2) &= \frac{1}{2}\rm{tr}\left |\hat{\rho}_1 - \hat{\rho}_2 \right |,
\end{align}
where $|\hat{M}|=\sqrt{\hat{M}^{\dagger}\hat{M}}$. For all quantum Markov processes, any two initial states will become less distinguishable during the time evolution. More precisely, the trace distance of any pair of initial states is a monotonically decreasing function of time, i.e., $D\left(\hat{\rho}_1\left(t+\Delta t\right),\hat{\rho}_2\left(t+\Delta t\right)\right)\leq D\left(\hat{\rho}_1\left(t\right),\hat{\rho}_2\left(t\right)\right)$ for $\Delta t>0$. Thus, a process can be defined to be non-Markovian if there exists a pair of initial states for which the trace distance  $D\left(\hat{\rho}_1\left(t\right),\hat{\rho}_2\left(t\right)\right)$ increases at some time $t$ of the evolution \cite{BLP2009, WildeQIBook}.      

\begin{figure}[h]
\centering
\includegraphics[width=1.0\columnwidth, draft=false]{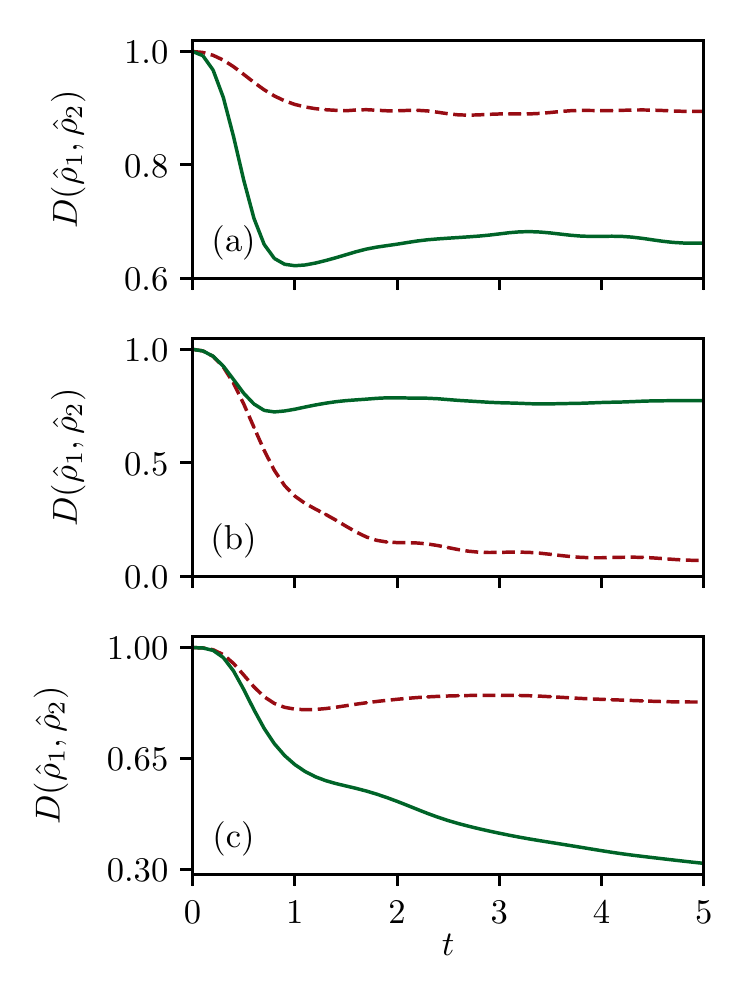}
\caption{\label{fig:NMwitness}Time evolution of trace distance of system’s density matrix. (a) Role of system-bath coupling strength:  The red dashed line represents smaller ($\gamma= 0.04$), and the green solid line represents larger ($\gamma=0.18$) system-bath coupling strength. Here, $U=1.125$. (b) The role of interaction $U$:  The red dashed line is for smaller interaction $(U=0.5)$, and the green solid line is for larger interaction $(U=1.4)$. Here, $\gamma=0.16$. In panels (a) and (b), $T_L=0.5, T_R=0.25$ and $\omega_c=1.5$. (c) The role of bath frequency cut-off $\omega_c$: The red dashed line is for lower cut-off $(\omega_c=1.5)$, and the green solid line is for higher cut-off $(\omega_c=1.75)$. Here, $T_L=0.125, T_R=0.0625$, $U=1.25$ and $\gamma=0.16$. }
\end{figure}

In Fig.~\ref{fig:NMwitness}, we look for signatures of non-Markovianity for various system and bath parameters. We use the trace distance $D(\hat{\rho}_1,\hat{\rho}_2)$ between two initial pure states of the system $\hat{\rho}_1=| \psi_1 \rangle_S \langle \psi_1 |$ and $\hat{\rho}_2=| \psi_2 \rangle_S \langle \psi_2 |$ with $|\psi_1 \rangle_S=|0 \rangle_S $ and $|\psi_2 \rangle_S = |2 \rangle_S$. Here, $|0\rangle_S$ represents the state with no bosons in the system, whereas for $|2\rangle_S$ the system has initially two bosons. A non-monotonous behavior for these two states would be a sufficient condition for the dynamics to be non-Markovian. The trace distance has the maximum value of $1$ for the two chosen initial states. It is clear from Fig.~\ref{fig:NMwitness}(a) that for weak coupling ($\gamma=0.04$) the dynamics does not show signatures of non-Markovianity, whereas increasing the coupling $\gamma$, the trace distance decreases non-monotonously resulting in non-Markovian dynamics. 
In Fig.~\ref{fig:NMwitness}(b), we focus on the effect of interactions, and we observe that stronger interactions (continuous green line), result in a slower dynamics which shows a more marked non-Markovian behavior (non-monotonicity of the distance) at shorter times. 
In Fig.~\ref{fig:NMwitness}(c), we consider the effect of bath frequency cut-off $w_c$. For smaller $w_c$, one expects a more marked non-Markovian dynamics, as the bath correlations decay more slowly \cite{OQSBook}. This is what we observe in Fig.~\ref{fig:NMwitness}(c) where the curve for $w_c=1.5$ (red dashed line) shows a more marked non-monotonous behavior than for $w_c=1.75$ (green solid line).

\section{Conclusions}
\label{sec:conclusions}
The interplay of interactions, temperature biases, and system-bath couplings may lead to rich and complex transport phenomena. In view of this, we investigate the nonequilibrium dynamics of an anharmonic oscillator coupled to two thermal baths using numerically exact TCMPS. First, we discuss the effectiveness of TCMPS to analyze the steady-state properties, especially compared to using the global master equation. We show that, for stronger interactions and/or weaker system-bath couplings, the GME represents the steady-state exactly,
however, it still cannot reproduce the short-time dynamics accurately. For stronger system-bath couplings or lower temperatures, the dynamics can only be reliably studied with TCMPS.
We have also shown that the average particle number, the particle and energy currents decrease as the interaction increases, and we have shown a non-linear dependence of the energy current versus system-bath coupling strength. We have also shown the emergence of non-Markovian dynamics in the strong coupling limit by measuring the trace distance. This non-Markovian behavior depends on the bath properties, on the system-bath coupling strength, but also on the magnitude of the interactions in the system. 

% In the future we plan to use the same tool to study more complex systems such as chains and structures in higher dimensions.
\tc{In the future, we plan to use the same tool to study quantum transport in more complex systems coupled to two heat baths as, for instance, higher-dimensional systems with frustration. The possibility to study strong coupling and beyond linear response can lead to effects, such as negative differential conductance \cite{BenentiRossini2009, BenentiZnidaric2009}. }

\begin{acknowledgments}
We thank B. K. Agarwalla, P. H\"anggi, A. Purkayastha, X. Xu and J. Thingna for fruitful discussions. 
D. P. acknowledges support from the Singapore Ministry of Education, Singapore Academic Research Fund Tier-II (Project No.~MOE2016-T2-1-065). C. G. acknowledges support from National Natural Science Foundation of China under Grant No. 11805279. The computational work for this article was partially performed on resources of the National Supercomputing Centre, Singapore (NSCC) \cite{NSCC}.
\end{acknowledgments}

\bibliographystyle{apsrev}

\begin{thebibliography}{100}
\bibitem{cotunnelingfootnote} \tc{We point out here that, from a numerical point of view, the study of very low temperatures or very high ones, within the TCMPS framework, can both lead to significant difficulties. In the very low temperature baths case, especially with weak system-bath coupling, the simulations time required for equilibration may be considerably large. For high temperatures, especially with large system-bath couplings, one may require a considerably large local Hilbert space and bond dimension to accurately study the dynamics for large enough times in order to observe a steady state. Thus, we restrict our conclusion within the parameters chosen in our paper.}
\end{thebibliography}

\appendix

\renewcommand{\thefigure}{A\arabic{figure}}
\setcounter{figure}{0}

\section{Steady-state occupation convergence}
\label{app:steadystateconvergence}

\begin{figure}[h]
\centering
\includegraphics[width=1.\columnwidth, draft=false]{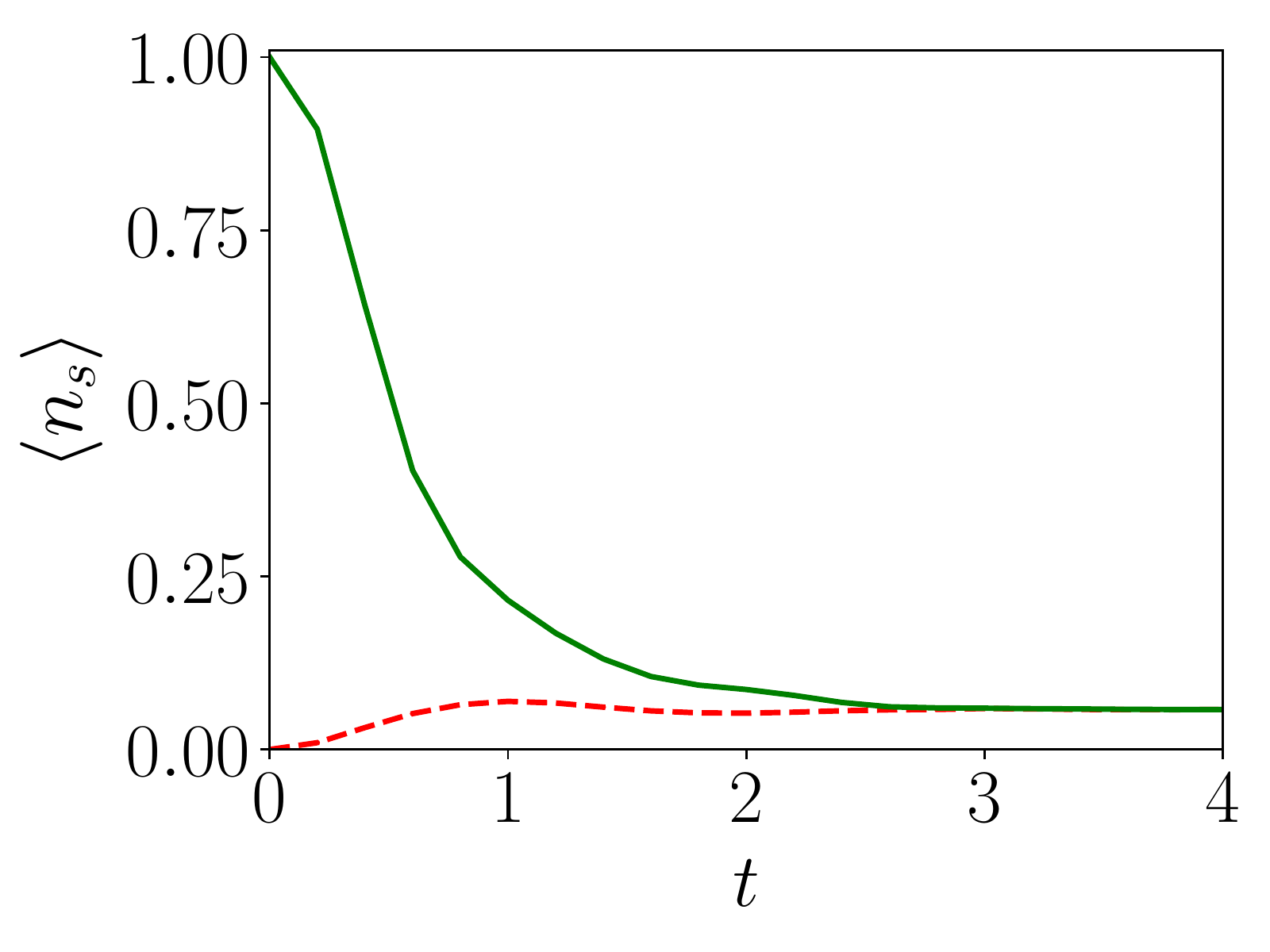}
\caption{\label{fig:steadystateconvergence} Time evolution of the system occupation $\langle {n}_S\rangle$ for different initial conditions: The green solid line is for one boson in the system $\left(| 1 \rangle \right)$ whereas the red dashed line is for an empty system at initial time $\left(|0\rangle \right)$. Other parameters used are as follows: $T_L=0.3125, T_R=0.125, \gamma=0.1225, U=1.5$, and $\omega_c=2.5$.}
\end{figure}

In Fig.~\ref{fig:steadystateconvergence}, we plot the system occupation as a function of time when they start initially with two distinct states in the systems: $|0\rangle_S$ and $| 1 \rangle_S$. It is clear from the figure that both curves converge, which indicates that our final states are, indeed, the steady states as they have the same occupation value from two different initial states.

\section{Global Master Equation (GME) approach}
\newcommand{\rhos}{\hat{\rho}_{\text{S}}}
\newcommand{\Aone}{\hat{A}_1\left(\omega\right)}
\newcommand{\Aonedag}{\hat{A}_1^{\dagger}\left(\omega\right)}
\newcommand{\Atwo}{\hat{A}_2\left(\omega\right)}
\newcommand{\Atwodag}{\hat{A}_2^{\dagger}\left(\omega\right)}
\label{app:GKSL}
We compare our TCMPS with a GME approach, where the effects of the bath are modeled by the Lindblad master equation,
\begin{align}
\label{eq:GLME}
    \frac{d \rhos}{dt}&=-\frac{{i}}{\hbar}\left[\mathcal{\hat{H}}_{\text{S}},\rhos\right]+\sum_{\nu=L,R}\mathcal{D}_\nu \left(\rhos\right),
\end{align}
where $\rhos$ is the system density matrix, $\nu=L,R$ is the index for the bath, and $\hbar$ is the Planck constant. $\mathcal{D}_L (\mathcal{D}_R)$ is the dissipator responsible for the coupling between the system and the left (right) heat bath, which acts globally on the system. The detailed expression of the dissipator is given as a Lindblad superoperator with the form
\begin{align}
    \mathcal{D}_\nu\left( \rhos \right) &=\sum_{0<\omega<\omega_c} \mathcal{J}\left(\omega\right)\left[1+n_\nu\left(\omega\right)\right]\bigg [\Aone \rhos \Aonedag \nonumber \\ \nonumber &- \frac{1}{2}\left\{\Aonedag\Aone,\rhos\right\}\bigg] \\ \nonumber
    &+\mathcal{J}\left(\omega\right)n_\nu\left(\omega\right)\bigg[\Atwo\rhos\Atwodag \\  &-\frac{1}{2}\left\{\Atwodag\Atwo,\rhos\right\}\bigg],
\end{align}
where $\omega=\epsilon_i-\epsilon_j$ is the energy difference of system eigenenergy with eigenstates $|i\rangle$ and $|j\rangle$. $n_\nu(\omega)=\left[\exp{\left(\hbar\omega/k_B T_\nu\right)}-1\right]^{-1}$ is the Bose-Einstein distribution for the heat bath on each side, and $k_B$ is the Boltzmann constant. $\mathcal{J}(\omega)=\gamma \omega$ is the ohmic spectral density for both baths, with cut-off frequency $\omega_c$. The Lindblad jump operators are
\begin{align}
    \Aone &= \sum_{\omega} |i\rangle\langle i|\as|j\rangle\langle j|, \\ 
    \Atwo &= \sum_{\omega} |i\rangle\langle i|\as^{\dagger}|j\rangle\langle j|,
\end{align}
which describe the transitions driven by the baths.

\onecolumngrid

\section{Derivation of system-bath energy current}
\label{app:heatcurrent}

The details of the left and right bath energy currents within the setting of TCMPS are discussed here. We start by obtaining the expression of energy current of the left and right baths from the original Hamiltonian from the system, the bath and interaction Hamiltonians. Although the discretized Hamiltonian is used here initially, in general, the bath oscillators form a quasicontinuum and could be rewritten as an integral. We will see this later when performing the chain-mapping technique. The total Hamiltonian is
\begin{align}
	\mathcal{\hat{H}}_{\text{S}} &= \frac{U}{2} n_{\text{S}} \left( n_{\text{S}} - 1 \right) + \mu n_{\text{S}}, \\ \nonumber
	\mathcal{\hat{H}}_{\text{B}}^{\nu} &= \sum_{k=1}^{N}\omega_k b_k^{\nu \dagger}b_k^{\nu}, \\ \nonumber
	\mathcal{\hat{H}}_{\text{I}}^{\nu} &=\sum_{k=1}^{N}\sqrt{\mathcal{J}_k}\left( a_{\mathrm{S}}^{\dagger}b_k^{\nu} + a_{\mathrm{S}} b_k^{\nu\dagger}\right),
\end{align}
where $\nu=L,R$ is the index for the left and right bath. $N$ is the number of modes in the bath. The energy current is defined by the continuity equation of the system Hamiltonian,
\begin{align}
\frac{\partial \left \langle \hat{\mathcal{H}}_{\text{S}}\right \rangle}{\partial t} + \nabla J_{\text{e}} &= 0,
\end{align}
where $J_{\text{e}}$ denotes the energy current. The above relation can be rewritten with respect to energy current from the left ($J_{\text{e}}^L$) and right ($J_{\text{e}}^R$) baths as
\begin{align}
\frac{\partial \left \langle\hat{\mathcal{H}}_{\text{S}}\right \rangle}{\partial t} &= \left \langle{i}\left [ \mathcal{\hat{H}}_{\text{tot}},\hat{\mathcal{H}}_{\text{S}}  \right]\right\rangle = - \left( J_{\text{e}}^R - J_{\text{e}}^L\right) = J_{\text{e}}^L - J_{\text{e}}^R,
\end{align}
with $\mathcal{\hat{H}}_{\text{tot}}=\mathcal{\hat{H}}_{\text{S}}+\mathcal{\hat{H}}_{\text{B}}^{L}+\mathcal{\hat{H}}_{\text{B}}^{R}+\mathcal{\hat{H}}_{\text{I}}^{L}+\mathcal{\hat{H}}_{\text{I}}^{R}$. We could easily obtain the expression for $J_{\text{e}}^L$ and $J_{\text{e}}^R$, respectively,
\begin{align}
\label{eq:originalheatcurrent}
	J_{\text{e}}^L &= 2U\sum_{k=1}^{N} \sqrt{\mathcal{J}_k} \mathrm{Im} \left \langle b_k^L \as^{\dagger}\ns \right \rangle +2\mu\sum_{k=1}^{N} \sqrt{\mathcal{J}_k} \mathrm{Im} \langle b_k^L \as^{\dagger} \rangle, \\ \nonumber
	J_{\text{e}}^R &= 2U\sum_{k=1}^{N} \sqrt{\mathcal{J}_k} \mathrm{Im} \left \langle \ns \as b_k^{R \dagger }  \right \rangle +2\mu\sum_{k=1}^{N} \sqrt{\mathcal{J}_k} \mathrm{Im} \langle \as b_k^{R \dagger}  \rangle,
\end{align}
where $\ns=\as^{\dagger} \as$ is the number operator of the system. Now, express the original bath annihilation and creation operators $b_k^{\nu}$ and $b_k^{\nu \dagger}$ in terms of the thermal Bogoliubov modes using Eq.~\eqref{eq:Bogliubovtransformation},
\begin{align}
b_k^{\nu} &= \cosh (\theta_k)a_{1,k}^{\nu}+\sinh (\theta_k)a_{2,k}^{\nu\dagger}, \\ \nonumber
b_k^{\nu \dagger} &= \cosh (\theta_k)a_{1,k}^{\nu\dagger}+\sinh (\theta_k)a_{2,k}^{\nu\dagger}.
\end{align}
Thus the energy current can be written as
\begin{align}
J_{\text{e}}^L &= 2U\sum_{k=1}^{N}g_{1,k}^L \mathrm{Im}\left \langle a_{1,k}^L\adagn \right \rangle +g_{2,k}^L \mathrm{Im}\left \langle a_{2,k}^{L\dagger}\adagn \right \rangle \\ \nonumber &+2\mu\sum_{k=1}^N g_{1,k}^L\mathrm{Im}\left \langle a_{1,k}^{L}\as^{\dagger} \right \rangle+g_{2,k}^L \mathrm{Im}\left \langle a_{2,k}^{L\dagger}\as^{\dagger} \right \rangle, \\ \nonumber
J_{\text{e}}^R &= 2U\sum_{k=1}^{N}g_{1,j}^R \mathrm{Im}\left \langle a_{1,k}^{R\dagger}\na \right \rangle +g_{2,k}^R \mathrm{Im}\left \langle a_{2,k}^R \na \right \rangle \\ \nonumber &+2 \mu \sum_{j=1}^N g_{1,k}^R\mathrm{Im}\left \langle a_{1,k}^{R\dagger}\as \right \rangle+g_{2,k}^R \mathrm{Im}\left \langle a_{2,k}^{R}\as \right \rangle,
\end{align}
with $g_{1,k}^{\nu}$ and $g_{2,k}^{\nu}$ previously defined as $g_{1,k}^{\nu}=\sqrt{\mathcal{J}_k}\cosh{(\theta_k^{\nu})}$, $g_{2,k}=\sqrt{\mathcal{J}_k}\sinh{(\theta_k^{\nu})}$ with $\cosh(\theta_k^{\nu})=\sqrt{1+n_{\nu}(\omega_k)}$, $\sinh(\theta_k^{\nu})=\sqrt{n_{\nu}(\omega_k)}$, and $n_{\nu}(\omega_k)=1/(e^{\omega_k/T_{\nu}}-1)$ in the main text ($\nu=L,R$).

Now, we perform the star-to-chain mapping. Before that, for the purpose of derivation, we rewrite the total discretized Hamiltonian $\mathcal{\hat{H}}_{\text{tot}}^{\text{dis}}$ after the thermofield transformation based on Eq.~\eqref{eq:spbhamiltonian}, and replace the summation $\sum_{k=1}^N(...)$ with the integral $\int dk (...)$ in the continuous limit as
\begin{align}
	\mathcal{\hat{H}}_{\text{tot}}^{\text{cont}} &= \mathcal{\hat{H}}_{\text{S}}+\sum_{\nu=L,R}\int_0^1dk\,k\left(a_{1,k}^{\nu \dagger}a_{1,k}^{\nu}-a_{2,k}^{\nu \dagger}a_{2,k}^{\nu}\right), \\ \nonumber
	&+ \sum_{\nu=L,R}\int_0^1 dk\, \Bigl[g_1^{\nu}(k)(\as^{\dagger}a_{1,k}^{\nu}+a_{1,k}^{\nu\dagger}\as) +g_2^{\nu}(k)(\as a_{2,k}^{\nu}+a_{2,k}^{\nu\dagger}\as^{\dagger}) \Bigl],
\end{align}
where $g_j^{\nu}(k)$ is the continuous counterpart of the coupling constant $g_{j,k}^{\nu}$, and the integration upper bound $1$ means that $\omega_k|_{k=1}=\omega_{N}$ which is the frequency cut-off of the heat bath.

Hence, the corresponding energy currents are as follows:
\begin{align}
\label{eq:Jheatafterthermofield}
J_{\text{e}}^L &= \underbrace{2U\int_0^1 dk\, g_1^L(k)\mathrm{Im}\left\langle a_{1,k}^L\adagn \right\rangle}_{\Delta_1}\\ \nonumber &+\underbrace{2U\int_0^1 dk\, g_2^L(k)\mathrm{Im}\left\langle a_{2,k}^{L\dagger}\adagn \right\rangle}_{\Delta_2}\\ \nonumber &+\underbrace{2\mu\int_0^1 dk \, g_1^L(k)\mathrm{Im}\left\langle a_{1,k}^L\as^{\dagger} \right\rangle+g_2^L(k)\mathrm{Im}\left\langle a_{2,k}^{L\dagger}\as^{\dagger} \right\rangle}_{\Delta_3}. \\ \nonumber
J_{\text{e}}^R &= 2U\int_0^1 dk\, g_1^R(k)\mathrm{Im}\left\langle a_{1,k}^{R\dagger}\na \right\rangle \\ \nonumber &+2U\int_0^1 dk\, g_2^R(k)\mathrm{Im}\left\langle a_{2,k}^R\na \right\rangle \\ \nonumber &+2\mu\int_0^1 dk \, g_1^R(k)\mathrm{Im}\left\langle a_{1,k}^{R\dagger}\as \right\rangle+g_2^R(k)\mathrm{Im}\left\langle a_{2,k}^R\as \right\rangle.
\end{align}
Now, we adopt the unitary transformation in Refs.~\cite{prior2010efficient,Plenio2010exactmapping} to transform the total system from a star configuration to a chain representation. The transformation is real and it leads to two new bosonic modes (for the simplicity of the expression, we neglect the superscript $\nu=L,R$ for baths),
\begin{align}
\label{eq:unitarytransformation}
a_{1,k}^{} &= \sum_{n=0}^{N_c-1}U_{1,n}^{}(k)d_{1,n+1}^{},\, a_{2,k}^{}=\sum_{n=0}^{N_c-1}U_{2,n}^{}(k)d_{2,n+1}^{},
\end{align}
where $U_{j,n}^{}(k)=g_j^{}(k)\pi_{j,n}^{}(k)/\rho_{n,j}^{}(j=1,2)$ and the $\{\pi_{j,n}^{}(k)\}$ series is monic orthogonal polynomials \cite{Plenio2010exactmapping,Plenio2014chainrepresentation, Plenio2011bookchapter} which obey

\begin{align}
\label{eq:MOPproperties}
\int_0^1 dk\,\mathcal{S}_j^{}(k)\pi_{j,n}^{}(k)\pi_{j,m}^{}(k) &=\rho_{n,j}^{2}\delta_{nm},
\end{align}
with newly defined spectral density in the transformed bath,
\begin{align}
\label{eq:newspectraldensity}
\mathcal{S}_1(k) &=g_1^{}(k)^2=\left(1+n(\omega_k)\right)\mathcal{J}_k, \\ \nonumber
\mathcal{S}_2(k) &=g_2^{}(k)^2=n(\omega_k)\mathcal{J}_k,
\end{align}
and $\int_0^1 dk\,\mathcal{S}_j(k)\pi_{j,n}^2(k) =\rho_{n,j}^2$, and $\pi_{j,0}(k)=1$. $N_c$ is the number of bosonic modes in the chain representation. We now move on to further express $J_{\text{h}}^L$ in terms of those new baths.
We first put back the bath indices $\nu$ and insert Eq.~\eqref{eq:unitarytransformation} into Eq.~\eqref{eq:Jheatafterthermofield}, and evaluate all three terms $(\Delta_1,\Delta_2,\Delta_3)$ from $J_{\text{h}}^L$,
\begin{align}
\label{eq:leftheatcurrentfourterms}
	\Delta_1 &= 2U\sum_{n=0}^{N_c-1}\int_0^1dk\, g_1^L(k)U_{1,n}^L(k)\mathrm{Im}\left \langle d_{1,n+1}^L\adagn \right \rangle \\ \nonumber
	&=2U\sum_{n=0}^{N_c-1}\int_0^1 dk \, \left[g_1^L(k)\right]^2\pi_{1,n}^L(k)/\rho_{n,1}^L \mathrm{Im}\left \langle d_{1,n+1}^L\adagn \right \rangle \\ \nonumber
	&=2U\sum_{n=0}^{N_c-1}\int_0^1 dk \, \mathcal{S}_1^L(k) \pi_{1,n}^L(k)/\rho_{n,1}^L \mathrm{Im}\left \langle d_{1,n+1}^L\adagn \right \rangle.
\end{align}
Now, multiply each side of Eq.~\eqref{eq:leftheatcurrentfourterms} by $\pi_{1,0}^L(k)$ and use Eq.~\eqref{eq:MOPproperties},
\begin{align}
\label{eq:delta1}
\Delta_1 &=2U\sum_{n=0}^{N_c-1}\int_0^1 dk \,\mathcal{S}_1^L(k) \pi_{1,n}^L(k)\pi_{1,0}^L(k)/\rho_{n,1}^L \mathrm{Im}\left \langle d_{1,n+1}^L\adagn \right \rangle \\ \nonumber
&=2U\sum_{n=0}^{N_c-1} \left(\rho_{n,1}^L\right)^2\delta_{n,0}/\rho_{n,1}^L \mathrm{Im}\left \langle d_{1,n+1}^L\adagn \right \rangle \\ \nonumber
&=2U \rho_{0,1}^L
\mathrm{Im}\left \langle d_{1,1}^L\adagn \right \rangle
\end{align}
where $(\rho_{0,1}^L)^2=\int_0^1 dk \, \mathcal{S}_1^L(k)\pi_{1,0}^{L}(k)^2=\int_0^1dk\, \mathcal{S}_1^L(k)$. Now, define the tunneling between the system and the first site in the bath chain as
\begin{align}
\beta_{1,0}^L&=\rho_{0,1}^L=\sqrt{\int_0^1dk\, \mathcal{S}_1^L(k)}.
\end{align}
Thus
\begin{align}
\Delta_1 &= 2U \beta_{1,0}^L
\mathrm{Im}\left \langle d_{1,1}^L\adagn \right \rangle.
\end{align}
For $\Delta_2$, it is also defined that $\beta_{2,0}^L=\rho_{0,2}^L=\sqrt{\int_0^1 dk \, \mathcal{S}_2^L(k)}$. Thus, we have
\begin{align}
\label{eq:delta2}
	\Delta_2 &= 2U \beta_{2,0}^L \mathrm{Im}\left \langle d_{2,1}^{L\dagger} \adagn \right \rangle.
\end{align}
Also, for $\Delta_3$,
\begin{align}
\label{eq:delta3}
	\Delta_3 &= 2\mu\left[\beta_{1,0}^L\mathrm{Im}\left \langle d_{1,1}^L \as^{\dagger} \right \rangle+\beta_{2,0}^L \mathrm{Im}\left \langle d_{2,1}^{L\dagger}\as^{\dagger} \right \rangle\right].
\end{align}
Add up all three terms, we arrive at the expression for $J_{\text{e}}^L$,
\begin{align}
\label{eq:finalleftheatcurrent}
J_{\text{e}}^L&= \Delta_1+\Delta_2+\Delta_3 \\ \nonumber
&=2U \beta_{1,0}^L
\mathrm{Im}\left \langle d_{1,1}^L\adagn \right \rangle  +2U \beta_{2,0}^L \mathrm{Im}\left \langle d_{2,1}^{L\dagger} \adagn \right \rangle \\ \nonumber
&+2\mu\left[\beta_{1,0}^L\mathrm{Im}\left \langle d_{1,1}^L \as^{\dagger} \right \rangle+\beta_{2,0}^L \mathrm{Im}\left \langle d_{2,1}^{L\dagger}\as^{\dagger} \right \rangle\right].
\end{align}
In a similar way, the expression for the energy current of the right bath $J_{\text{e}}^R$ can be obtained as
\begin{align}
\label{eq:finalrightheatcurrent}
J_{\text{e}}^R&= 2U \beta_{1,0}^R \mathrm{Im}\left \langle d_{1,1}^{R\dagger} \na\right \rangle+2U\beta_{2,0}^R\mathrm{Im}\left \langle d_{2,1}^R \na \right \rangle \\ \nonumber
&+ 2\mu\left[\beta_{1,0}^R\mathrm{Im}\left \langle d_{1,1}^{R\dagger}\as \right \rangle+\beta_{2,0}^R \mathrm{Im}\left \langle d_{2,1}^R \as \right \rangle \right].
\end{align}
\\
\\
\twocolumngrid
\section{Derivation of system-bath particle current}
\label{app:particlecurrentderivation}

The expression for particle current between the system and the bath is obtained via the continuity equation of the system occupation $n_{\text{S}}$,
\begin{align}
\frac{\partial \left \langle n_{\text{S}}\right\rangle}{\partial t} + \nabla J_{\text{p}} &=0,
\end{align}
where $J_{\text{p}}$ denotes the particle current, and $n_{\text{S}}=a_{\text{S}}^{\dagger}a_{\text{S}}$ is the system occupation. Similar to that of the energy current, we can rewrite this relation in terms of particle current from the left ($J_{\text{p}}^L$) and the right bath ($J_{\text{p}}^R$) as
\begin{align}
	\frac{\partial \left \langle n_{\text{S}}\right\rangle}{\partial t} &=\left\langle {i}\left[ \mathcal{\hat{H}}_{\text{tot}}, n_{\text{S}} \right]\right\rangle=-\left(J_{\text{p}}^{R} - J_{\text{p}}^L \right) = J_{\text{p}}^L - J_{\text{p}}^R.
\end{align}
Following a similar derivation as in Appendix.~\ref{app:heatcurrent}, we  obtain the expression for both particle currents as
\begin{align}
	J_{\text{p}}^L &= 2\beta_{1,0}^L \operatorname{Im}\left \langle a_{S}^{\dagger} d_{1,1}^L \right \rangle + 2\beta_{2,0}^L \operatorname{Im}\left \langle d_{2,1}^{L \dagger} a_{\text{S}}^{\dagger}\right \rangle.  \\ \nonumber
	J_{\text{p}}^R &= 2\beta_{1,0}^R \operatorname{Im} \left \langle a_{\text{S}}d_{1,1}^{R\dagger} \right \rangle + 2\beta_{2,0}^R \operatorname{Im}\left \langle a_{\text{S}}d_{2,1}^R \right \rangle.
\end{align}

\end{document}